\documentclass[%
 aip,
 sd,%
 amsmath,amssymb,
 reprint,%
]{revtex4-1}

\usepackage{amsmath}
\usepackage{amssymb}
\usepackage{graphicx}
\usepackage{color}
\usepackage{verbatim}

\graphicspath{{figures/}{}}
\begin{document}

\title{Evidence for universal relationship between the measured $1/f$ permittivity noise and loss tangent created by tunneling atoms}


\author{A. N. Ramanayaka}
\affiliation{Laboratory for Physical Sciences, College Park, MD 20740, USA}
\affiliation{Department of Physics, University of Maryland, College Park, MD 20742, USA}
\author{B. Sarabi}
\affiliation{Laboratory for Physical Sciences, College Park, MD 20740, USA}
\affiliation{Department of Physics, University of Maryland, College Park, MD 20742, USA}
\author{K. D. Osborn}
\affiliation{Laboratory for Physical Sciences, College Park, MD 20740, USA}
\affiliation{Joint Quantum Institute, University of Maryland, College Park, MD 20742, USA}


\date{\today}

\begin{abstract}
Noise from atomic tunneling systems (TSs) limit the performance of various resonant devices, ranging in application from astronomy detectors to quantum computing. Using devices containing these TSs, we study the $1/f$ permittivity noise and loss tangent in two film types containing different TS densities. The noise reveals an intrinsic value as well as the crossover to power-saturated noise. The intrinsic $1/f$ noise fits to the temperature dependence $1/T^{1+\mu}$ where $0.2\le\mu\le0.7$, which is related to previous studies and strongly interacting TS. An analysis of the noise normalized by the loss tangent and temperature is quantitatively identical for two film types, despite a factor of 5 difference in their loss tangent. Following from the broad applicability of the TS model, the data supports a universal relationship for amorphous-solid produced permittivity noise. The quantity of the observed noise particularly supports a recent model in which noise is created by weak TS-TS interactions.
\end{abstract}



\maketitle



Noise from atomic tunneling systems (TSs) is believed to be responsible for performance limitations in various resonant devices, from kinetic inductance detectors in astronomy to measurement and bus resonators in quantum computing. TS noise is also likely related to qubits in quantum computing since they can create charge noise in a qubit \cite{Nakamura1} and critical current noise in a Josephson junction\cite{Nugroho}. While TSs in amorphous solids have been known for decades, their rediscovery in superconducting qubits \cite{Simmonds1, Martinis1} has led to recently increased interest. This has led to many fundamental studies of individual TS in qubits \cite{Simmonds1, Cole1, Lisenfeld1, Lisenfeld2, Lisenfeld3, Pappas1, Stoutimore1, Zaretskey1, Kim1,Sarabi1}, and microwave loss tangent and internal quality factor measurements to lower the loss in structures\cite{Devoret1, Khalil2, Paik1, OConnell1, Khalil3}. After the initial qubit decoherence studies, the performance-limiting 1/f frequency noise in resonators was discovered. This noise was found to decrease with increasing field amplitude (or power) and was linked to TS due to the calculated contribution of TS in different resonator geometries \cite{Gao1} . 

While the loss phenomena and some coherent dynamics of TSs is understood in terms of a standard model of two-level systems \cite{Phillips1,Anderson1}, the dynamics of TS that lead to frequency noise is an unresolved issue. In the standard model, the loss tangent saturates (decreases) below its intrinsic value of $\tan\delta=\pi P_0 p^2 / 3 \epsilon$ as the ac drive field is increased, where $P_0$ is the density of TLS, $p$ is the dipole moment, and $\epsilon$ is the permittivity as the Rabi frequency of resonant TS exceeds a decoherence rate. An early theoretical studies of the TS $1/f$ noise for qubits, using the standard distribution of TSs, found a negligible amount of noise saturation with increasing field strength \cite{Yu1}. A semi-empirical model for noise was later proposed which included the observed noise saturation \cite{Gao2}. This was followed by a theoretical model which showed that deterministic dynamics from TS can squeeze noise, creating enhanced frequency fluctuations, but the study only included a simplified model of TS \cite{Takei1}. More recently, an experimental study uncovered interesting phenomena: the $1/f$ noise was found to decrease as the temperature increased with a particular power law \cite{Brunett1}. A theoretical model mentioned \cite{Brunett1, Faoro1} was proposed in which strong interactions between TS create the observed frequency noise, but it also used a nonstandard energy distribution of TS. More recently, a weak TS-TS interaction model of the noise was proposed based on the standard distribution of TSs \cite{Burin1}. In the previous two noise models $\tan\delta$ is predicted to exhibit a linear-response regime of noise and a particular saturation (or decrease) in noise above a crossover field. To the best of our knowledge this crossover has not been observed, and a quantitative relationship between the noise magnitude and the loss from TS in a material has not been established. 

Here we report on the TS-induced frequency noise from resonators, all with resonance frequencies of approximately $\nu_0 = 4.7 $GHz, along with their dielectric loss tangents. The frequency noise spectral density $S_{\delta\nu/\nu}(f)$ is created by TSs and is $1/f$ (1/frequency) type. The microwave power dependence of this noise shows an intrinsic (unsaturated) noise at low powers and a saturation at higher powers which is compared to the loss tangent saturation character. Furthermore, we find that the temperature dependence of the intrinsic noise level has an interesting increase as the temperature decreases, which is compared to previous measurements of 1/f noise \cite{Kumar1, Brunett1} and a strongly coupled TS-TS noise theory \cite{Faoro1}. One of the main findings of this work is that in the unsaturated regime the $1/f$ permittivity noise power divided by the loss tangent is approximately the same for amorphous films with significantly different loss tangents and supports a universal relationship for amorphous solids in general. This comparison and the quantitative value of the noise supports the noise model of Ref. [\onlinecite{Burin1}], with weakly coupled TS in a standard distribution, which cause coherent TS dephasing as a result of thermally excited TS at low energy.


\begin{figure}
\includegraphics[width=0.483\textwidth]{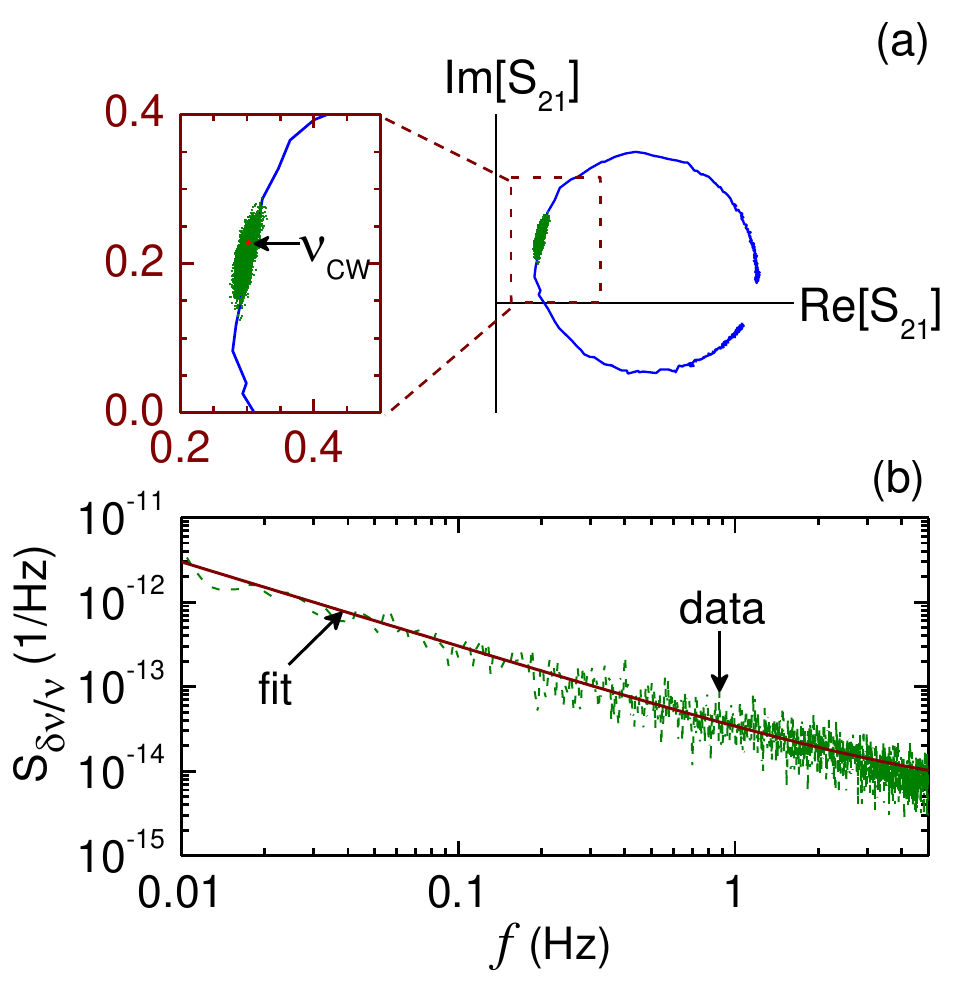}
\caption{(Color online) (a) Im$[$S$_{21}]$ versus Re$[$S$_{21}]$ for both swept- (line) and CW- (dots) frequency data, where the CW frequency data show frequency fluctuations along to the resonance circle of the swept-frequency data.  (b) Fractional frequency noise spectral density $S_{\delta \nu/\nu}$ at temperature $15$ mK and input microwave power $P_{in}=-119.5$ dBm. Solid line is a fit to $A/[f/(1$Hz$)]+B$, where $A$ gives the TS produced $1/f$ noise.} 
\label{figure1}
\end{figure}

Microwave LC resonators were fabricated from superconducting aluminum (Al) films with a $d=250$ nm thick film of amorphous (a-)Si$_3$N$_4$  forming the capacitor dielectric. Here we will report on measurements on two film types with substantially different loss tangents, which are denoted by Type-1 and Type-2, whose growth are described in Ref. [\onlinecite{Paik1}]. The capacitances $C$ were either 1.5 or 4.6 pF, and are comparable to many microwave resonators, where details of the device structure are given in Ref.[\onlinecite{Sarabi1}]. The dielectric volume ($V$) of the TS-hosting amorphous- (a-)Si$_3$N$_4$ film within $C$ is $5000$ $\mu$m$^{3}$. The trilayer geometry of the capacitor provides a $100\%$ participation of the material and results in a simple relationship between the loss tangent $\tan\delta=1/Q_{i}$ and the internal quality factor $Q_{i}$, and additionally the permittivity noise $S_{\delta\epsilon/\epsilon}=4S_{\delta\nu/\nu}$ and the frequency noise ($S_{\delta\nu/\nu}$). 

Measurements were carried out in a cryogen-free dilution refrigerator. The input microwave line is attenuated by approximately 75 dB at different stages in the refrigerator to prevent a single thermal photon excitation in the resonator. The output microwave line is connected to a cryogenic high-electron-mobility-transistor (HEMT) low-noise amplifier mounted at 3 K, and the device is isolated from it by multiple cryogenic circulators.

\begin{figure}
\includegraphics[width=0.483\textwidth]{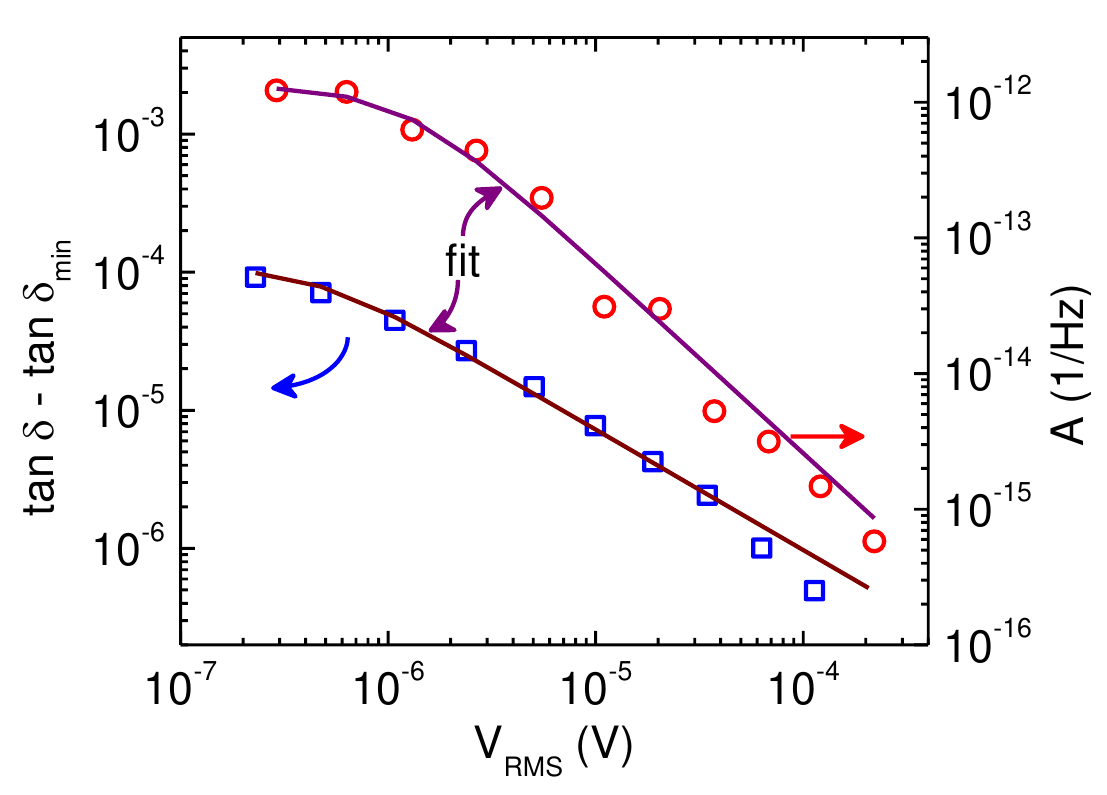}
\caption{(Color online) Loss tangent, $\tan\delta_{0}-\tan\delta_{min}$, ($\square$, left-axis), and fit extracted $1/f$ noise component, $A$, ($\bigcirc$, right-axis), as a function of RMS voltage ($V_{RMS}$) across the trilayer capacitor. $\tan\delta_{min}$ was measured at high fields and is not attributable to TS. Similar to the loss tangent, the onset of $1/f$ noise power saturation is clearly observed, such that an intrinsic region of noise is identified (see text for details).}
\label{figure2}
\end{figure}


\begin{figure}
\includegraphics[width=0.483\textwidth]{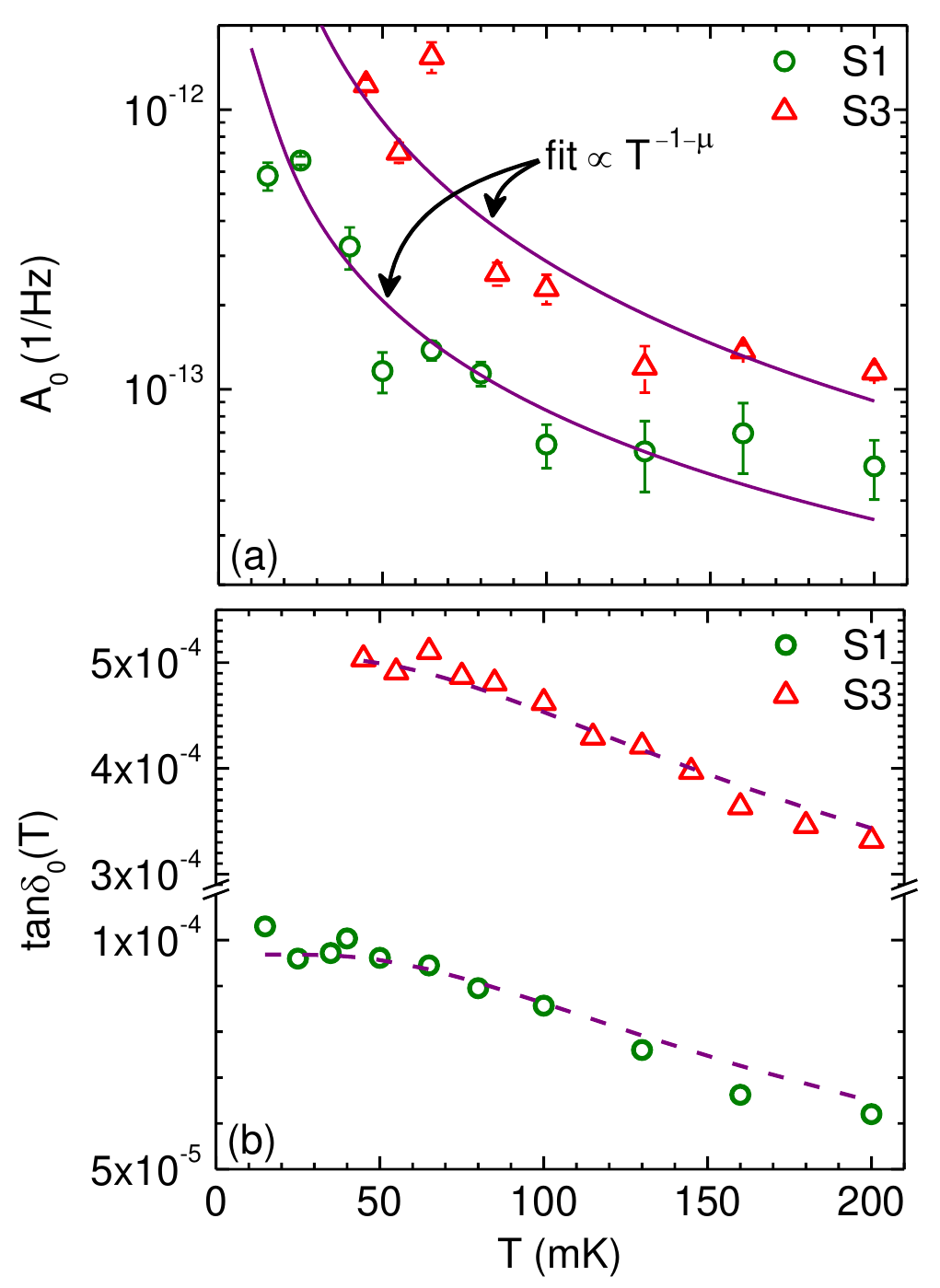}
\caption{(Color online) (a) Temperature dependence of low-power $1/f$ noise $A_{0}$ shown for representative samples: S1 of type 1 ($\bigcirc$) and S3 of type 2 ($\bigtriangleup$). Here the error bars indicate the standard deviation of the noise values determined from fitting. The solid line is a numerical fit of $A_{0}$ for samples S1 and S3 to the temperature dependent function shown, where $\mu = 0.30 \pm 0.17$ and $0.65 \pm 0.17$, respectively. (b) Temperature dependence of low-power loss tangent $\tan\delta_{0}$ for samples S1 ($\bigcirc$) and S3 ($\bigtriangleup$). Sample type-1 (S1) has a loss tangent,$\tan\delta_{0}(T<60 $mK$)$, which is approximately $5$ times smaller than that of type-2 (S3).}
\label{figure3}
\end{figure}

Microwave transmission (S$_{21}$) measurements on the devices were performed with the VNA on swept frequency mode to characterize the loss tangent \cite{Khalil2}. The films in the devices have unsaturated loss tangents of either $1.0 \times 10^{-4}$ (Type-1) or $5.0 \times 10^{-4}$ (Type-2).  In a subsequent measurement with the VNA set to CW mode, S$_{21}$ was measured as a function of time while keeping the CW frequency fixed near $\nu_0$. Figure \ref{figure1}(a) shows the real and imaginary parts of S$_{21}$ from swept- and CW- frequency measurements made at $T=15$ mK an input microwave power $P_{in}=-119.5$ dBm. With the input frequency fixed, both the noise quadratures (Re[S$_{21}$] and Im[S$_{21}$]) measure a much larger noise on resonance compared to frequencies far from resonance. The fluctuations along the transmission circle are a consequence of especially enhanced fluctuations in the resonance frequency, and the phase of S$_{21}$ is used to determine the resonance frequency in time, using the phase of  swept-frequency measurements obtained at the same power and temperature.

\begin{figure}
\includegraphics[width=0.483\textwidth]{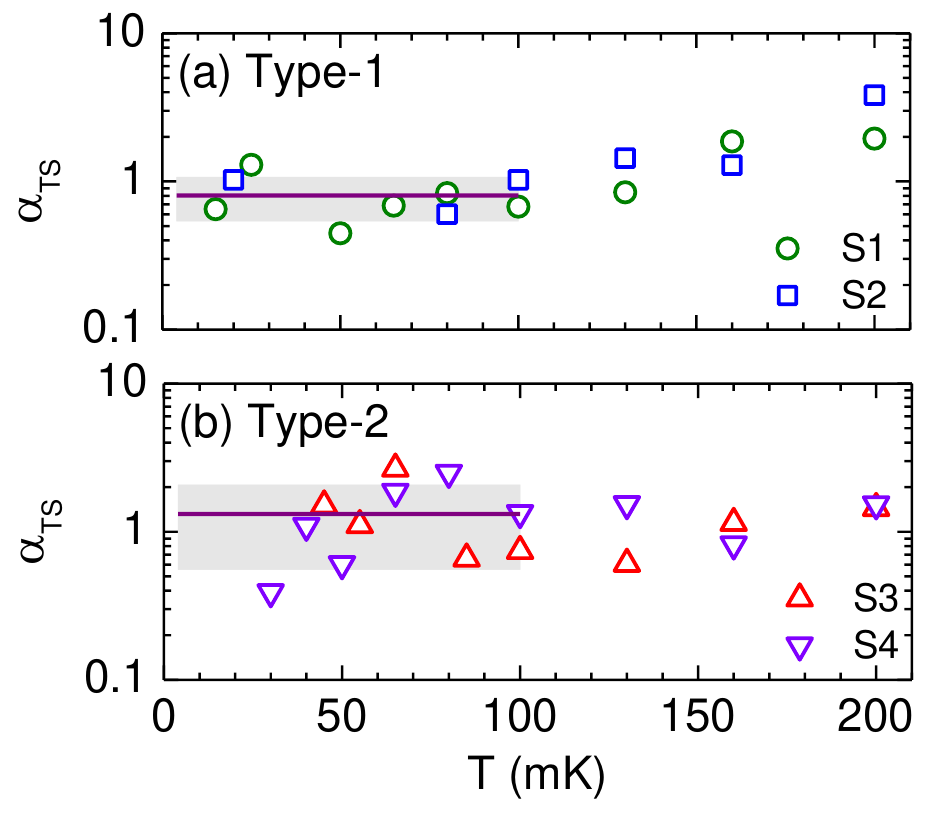}
\caption{(Color online) The scaled $1/f$ noise, $\alpha_{TS}$, which is proportional to the noise divided by the loss tangent and temperature, is shown for two a-Si$_3$N$_4$ film types: (a) Type-1 samples S1 ($\bigcirc$) and S2 ($\square$) and (b) Type-2 samples S3 ($\bigtriangleup$) and S4 ($\bigtriangledown$). For these amorphous films with very different TS densities a fit to $\alpha_{TS}$ for $T<110$mK$\approx\hbar\omega/k_B$ shows the same value within experimental uncertainty which is discussed in terms of universality of films and the noise model in the text.}
\label{figure4}
\end{figure}

$S_{\delta\nu/\nu}$, is extracted from time-series data using Welch's method \cite{Welch1}, and $1/f$-type behavior is observed at low noise frequencies ($f \le 1$ Hz), see Fig.\ref{figure1}(b). Furthermore, on resonance noise power is approximately two orders of magnitude higher than the off resonance value. For further analysis, $S_{\delta \nu/\nu}$ was fitted to a model $S_{\delta \nu/\nu} (P_{in} , T)=A/[f/(1$Hz$)]+B$, where $A$ and $B$ represents $1/f$ and white noise components, respectively. Unlike $A$, $B$ decreases monotonically with increasing $P_{in}$ and does not change with temperature, as it is only dependent on the measurement system noise (not shown). 

The input power $P_{in}=V_{in,RMS}^2/Z_0$ dependence of loss tangent $\tan\delta$ at $T=10$ mK is shown in Fig.\ref{figure2}. The root mean square (RMS) of the voltage across the capacitor plates $V_{RMS}=V_{in,RMS}\times Q / \sqrt{\pi Z_{0} C \nu_{0} Q_{c}}$ due to ac electric field across the trilayer capacitor $C$, can be expressed in terms $P_{in}$, the characteristic impedance $Z_0$, and the inverse total quality factor, $1/Q=1/Q_{c}+1/Q_{i}$. The coupling quality factor $Q_{c}$ is approximately $2.3 \times 10^4$ and $5.2 \times 10^3$ for film type 1 and 2, respectively, as obtained from swept-frequency S$_{21}$ data \cite{Khalil1}.  

A decrease in $\tan\delta$ with increasing input microwave power can clearly be seen due to power saturation of TSs at higher input powers. In both film types the loss tangent reaches a minimum $\tan\delta$, $\tan\delta_{min}$, which is practically independent of temperature up to $200$ mK near the highest measured input powers, where the photon number is $\ge 10^{6}$, and not caused by TS. The average value of $\tan\delta_{min}$ is used as a fixed parameter in the fit to the loss tangent $\tan\delta=\tan\delta_{0}/[1+(V_{RMS}/V_{L})^2]^{\beta_{L}} + \tan\delta_{min}$, where $\tan\delta_{0}(T)$ is the low-power loss tangent and $V_{L}$ is the crossover voltage of power saturation. In the standard TS theory, $\beta_{L}\approx0.5$ and $E_{L}=V_{L}/d=\sqrt{3}\hbar/2p\sqrt{T_{1,min}T_{2}}$, where $T_{1,min}$ is the minimum relaxation time for a TS on resonance, $T_{2}$ is the coherence time. Using a nonlinear least squares fit to the data, we extract the $\beta_{L}=0.43\pm0.03$ in sample S4, which has the smallest uncertainty in fitting. A fit to the standard theory of loss using a range similar to data gives $\beta_{L}=0.48$, which is only slightly larger than the experimental values. 

As shown in Fig. 2, the $1/f$ noise lowers (saturates) with increasing powers at high powers similar to $\tan\delta$ \cite{Kumar1, Gao1, Gao2, Brunett1, Faoro1}. Interestingly, at low power, and similar to the loss tangent, we find that $A$ reaches a power-independent (intrinsic) noise level, which to the best of our knowledge has not been previously observed. Similar to $\tan\delta$, we fit $A(V_{RMS})$ to a model of the form $A=A_{0}/[1+(V_{RMS}/V_{N})^2]^{\beta_{N}}$, where $A_{0}$ is the temperature dependent unsaturated $1/f$ noise and $V_{N}$ is the noise crossover voltage. The fit extracted values are $A_{0}=(1.3\pm0.1)\times10^{-12}$ Hz$^{-1}$, $V_{N}=(1.0\pm0.4)\times10^{-6}$ V$_{RMS}$, and $\beta_{N}=0.7\pm0.2$. The experimental result, $\beta_{N}=0.7\pm0.2$, shows a slightly larger slope than loss saturation, $\beta_{L}=0.4\pm0.07$. Furthermore, our value of $\beta_{N}$ agrees better with the weak interaction model of Ref. [\onlinecite{Burin1}], which gives $\beta_{N}=0.67$ for our data range, than with the strong interaction model of Ref. [\onlinecite{Faoro1}], where $\beta_{N}=0.5$.  

The unsaturated $1/f$ noise $A_{0}(T)$ is plotted in Fig.\ref{figure3}(a) for two samples of different type, and show a similar temperature dependence [see Fig.\ref{figure3}(a)]. The values of $A_{0}$ are fitted to a function $A_{0} \propto T^{-1-\mu}$, and the fit extracted values of $\mu$ for the devices S1 and S3 are $0.30 \pm 0.17$ and $0.65 \pm 0.17$, respectively. A similar temperature dependence from TS was observed in resonators previously\cite{Brunett1} with $\mu\simeq0.4$, which was used to support a strong TS-TS interaction model \cite{Brunett1, Faoro1}, but this behavior is also similar in the recent model with weak interactions such that we do not analyze it further. Similar to loss in strongly coupled loss model \cite{Faoro2}, the number of strongly coupled TS must be substantial. Using the strong interaction noise model with our calibrated quantity of permittivity noise gives the number of strongly-interacting TS $N_{TS}\simeq10^{-4}$ four orders of magnitude smaller than required in the model.

In Fig.\ref{figure3}(b), we plot low-power loss tangent $\tan\delta_{0}(T)$ for both film types, represented by S1 and S3, versus the bath temperature $T$. From the plots it is clear that the low temperature loss tangents, mentioned earlier, differ by a factor of 5. As mentioned above, the TS loss is determined from the measured loss minus an experimentally measured minimum $\tan\delta_{min}$, for each sample, such that the standard model of loss saturation is found, e.g. in Fig.\ref{figure2}, over a range of two decades. The loss is measured only for $T<200$mK to avoid affects from quasiparticles. The thermal saturation of tan$\delta_{0}(T)$ can agree with the standard model tan$ \delta \propto \tanh(\hbar \omega/ 2 k_B T)$, if some TS remain at a temperature of 200 mK throughout the temperature sweep, and in general this is may happed due to thermal boundaries at the aluminum film \cite{ThermalImpedance} which separates our dielectric film from the substrate. Although a correction with this assumption can be made, we choose to work with the data without any model dependence. This weakened temperature dependence is shown by comparing the data with a function derived from nonstandard assumptions of the initial state, $\tan \delta \propto 1- \exp(-\hbar \omega/ k_B T)$ \cite[The thermal saturation theory is not verified or used below in the analysis]{Vural1}.

In the weak TS-TS interaction model \cite{Burin1}, the low temperatures ($k_{B}T\ll \hbar\omega$) noise is created by spectral diffusion \cite{Black1} of the total permittivity coupled to low energy TS that are in thermally excited. To conveniently parameterize our noise we adopt the quantity $\alpha_{TS}=A_{0}\times N_{th} / \tan^{2}\delta$, where $N_{th}=P_{0}Vk_{B}T$ is the number of thermally activated TS, $\tan\delta$ is the measured loss tangent, and $A_{0}$ is the $1/f$ noise component taken at $1$ Hz. For the unsaturated regime we use $P_{0}=3 \tan \delta_0 / \pi p^2$, and approximate the distribution with a single moment at $p=6$ Debye for both film types. Since $P_{0}\propto\tan\delta_0$ this scaled noise is proportional to the experimental noise divided by the loss tangent and temperature at the lowest temperatures. This experimentally determined quantity is the same for both film types, as shown in the panels to Fig.\ref{figure4}, where the average values for $T\le100$ mK in Type-1 and Type-2 films are $0.80\pm0.26$ and $1.32\pm0.76$, respectively. Since the normalized noise, $\alpha_{TS}$, is nearly the same between materials even though the normalization factor from loss tangent differs by a factor of 5, this shows a convincing demonstration of how noise scales linearly with loss tangent in an amorphous film. Since amorphous films are believed to all follow the same model of TS, up the the value of the loss tangent or TS density, we expect this relationship to hold universally (generally) for any amorphous film with measurable TS. Both films show a value consistent with  $\alpha_{TS}\simeq1.0$ and this value is in quantitative agreement with the the weak interaction noise according to the parameters of Ref. \onlinecite{Burin1}. 

Our resonators have a high quality factor and can be compared to coplanar resonators where the TS are on surfaces, where for example, the low-power  $Q_i=10^6$ can be due to a loss tangent of $5.0 \times 10^{-4}$ and a dielectric participation of $0.2\%$. It is not surprising that the large slope $\beta_{N}=0.7\pm0.2$ in our device was not seen in earlier devices, since the TS are spread out in field strengths, which has the qualitative effect of reducing the power dependence of $S_{\delta\nu/\nu}$ relative to that in our film where the TS all experience the same saturation crossover due to a homogeneous field. Since we see a temperature dependence similar to previous resonators \cite{Brunett1} , we believe it likely that high-Q factor superconducting resonators in general have a their frequency noise also created by weak TS-TS interactions. Our resonators are made of typical amorphous films, and work spanning the previous four decades has established that the standard model generally describes the main dielectric, thermal and acoustic properties of these solids,  e.g. Ref. \onlinecite{Phillips1, Anderson1, Hunklinger1, PohlRev}. As such we think it is clear that an analysis of the resonant frequency noise with loss data in amorphous films is necessary and possibly sufficient to uncover a standard model of noise. 

In conclusion, we have measured the $1/f$ frequency noise of superconducting resonators, limited by atomic tunneling systems (TS) and related to devices in astronomy and quantum computing. Interestingly, the $1/f$ noise $S_{\delta\nu/\nu}$ varies as $T^{-1-\mu}$ where $\mu$ is 0.3 to 0.7, suggesting that our measured phenomenon is related to previous data used to support a strong interaction noise theory. Microwave power dependent studies suggest that the $1/f$ noise has the form $A_{0}/[1+(V_{RMS}/V_{N})^{2}]^{\beta_{N}}$, where $\beta_{N}=0.7\pm0.2$, indicating the existence of an intrinsic noise level $A_{0}$ and a larger saturation slope than previously observed. A quantitative analysis of the 1/f noise was performed by scaling the noise by the measured loss tangent and temperature. This forms a quantity which is unchanged with film type at low temperatures, despite using a loss tangent that differs by a factor of 5. From the data and the nature of amorphous solids we expect the scaled noise to represent a universal parameter. The scaled noise data supports a recent weak interaction noise model. Finally, since noise is related to loss this study implies that $1/f$ noise can be predicted from loss for a given device geometry. 


The authors would like to thank A. L. Burin, L. Faoro, D. C. Vural, Y. Rosen, S. Gladchenko, F. C. Wellstood, and R. P. Budoyo for helpful discussions.

\bibliography{references}

\end{document}